\documentclass[twocolumn,preprintnumbers,aps,pra,showpacs]{revtex4}
\usepackage{graphicx}

\begin{document}

\title{Phase-stable source of polarization-entangled photons using a polarization Sagnac interferometer}

\author{Taehyun Kim,\footnote{Electronic address: thkim@mit.edu}
Marco Fiorentino\footnote{Now at Hewlett-Packard Laboratories, 1501 Page Mill Road, Palo Alto, CA 90304, USA.}, and Franco N. C. Wong}

\affiliation{Research Laboratory of Electronics, Massachusetts Institute of 
Technology, Cambridge, Massachusetts 02139, USA}
\date{\today}

\begin{abstract}
We demonstrate a simple, robust, and ultrabright parametric down-conversion source of polarization-entangled photons based on a polarization Sagnac interferometer.  Bidirectional pumping in type-II phase-matched periodically poled KTiOPO$_4$ yields a measured flux of 5\,000 polarization-entangled pairs/s per mW of pump power in a 1-nm bandwidth at 96.8\% quantum-interference visibility. The common-path arrangement of the Sagnac interferometer eliminates the need for phase stabilization for the biphoton output state.   
\end{abstract}


\maketitle

Polarization-entangled photons are essential quantum resources for many applications in quantum information processing, including quantum cryptography \cite{QCrypt}, teleportation \cite{NJP}, and linear optics quantum computing \cite{KLM}.  The standard technique for efficiently producing polarization entanglement is by means of spontaneous parametric down-conversion (SPDC) in a $\chi^{(2)}$ nonlinear crystal such as beta barium borate (BBO) or periodically poled KTiOPO$_4$ (PPKTP).  In SPDC a pump photon is converted into two subharmonic photons, and the SPDC outputs can be arranged in various configurations to yield polarization entanglement between the photon pair.  For most practical applications, it is desirable to have a high flux of entangled photon pairs for a given spectral bandwidth, together with a high degree of entanglement.   One can quantify the performance of a down-conversion source in terms of its spectral brightness, namely the detected pairs/s/mW of pump power per nm of optical bandwidth, and its quantum-interference visibility.

A common approach uses a thin BBO crystal under type-II phase matching to generate non-collinearly propagating photon pairs that are polarization entangled \cite{NoncollinearEntangle}.  This simple arrangement requires a small aperture to restrict the field of view in order to obtain a high degree of entanglement, and consequently the flux is generally low.  A slightly different approach using a long PPKTP crystal with collinearly propagating outputs yields a higher spectral brightness of 300 pairs/s/mW/nm after postselection with a  50-50 beam splitter \cite{SingleBeamQI}. The increased flux is the result of a longer crystal and more efficient pair collection with the use of collinear propagation.  Yet, because of spatial-mode distinguishability in both approaches, apertures must be used and most of the output photon pairs are not collected.  

We have recently demonstrated a bidirectionally pumped down-converter that eliminates the constraint of spatial-mode distinguishability and obtained a detected flux of $\sim$12\,000 pairs/s/mW in a 3-nm bandwidth with a quantum-interference visibility of 90\% \cite{MZIBiPump}.  In this bidirectional pumping geometry, we used a single PPKTP crystal to implement a configuration of two coherently-driven SPDC sources whose outputs  were combined interferometrically with a Mach-Zehnder (MZ) interferometer.  The output photon pairs are polarization entangled over  the entire spatial cone and spectrum of emission without the need for spatial, spectral, or temporal filtering, thus allowing a much larger fraction of the output to be collected.  A similar implementation for nondegenerate polarization entanglement using a more efficient periodically poled lithium niobate (PPLN) crystal yielded a fiber-coupled generation rate of 300 pairs/s/mW in a narrow bandwidth of 50 GHz \cite{PPLNBiPump}.  The MZ source of Ref.~\cite{MZIBiPump} has two limitations.  The first problem is that the MZ setup is sensitive to environmental perturbation and hence it requires active servo control of a pump interferometer in order to set the phase of the biphoton output state.  The second problem is that the apparatus suffers from spatial-mode clipping in the arms of the MZ interferometer that reduces the quality of its entanglement.  

Shi and Tomita have recently used a Sagnac interferometer and type-I down-conversion for generating pulsed polarization entanglement in which its immunity to perturbations is noted \cite{Sagnac1}.  This Sagnac source used a beam splitter to separate the output photon pairs with a 50\% postselection success rate and a quantum-interference visibility of 71\% was achieved.  In this work we demonstrate a new version of the bidirectionally pumped down-conversion source that utilizes a polarization Sagnac interferometer (PSI).  Our new source uses type-II down-conversion and a polarization beam splitter for combining the outputs such that no postselection is required and it eliminates the need for active phase control.  Moreover, our polarization Sagnac source is implemented in a small package that yields a higher spectral brightness with a high degree of entanglement: $\sim$5\,000 detected pairs/s/mW/nm at a 96.8\% visibility.

\begin{figure}[b]
  \begin{center}
    \includegraphics[width=8cm]{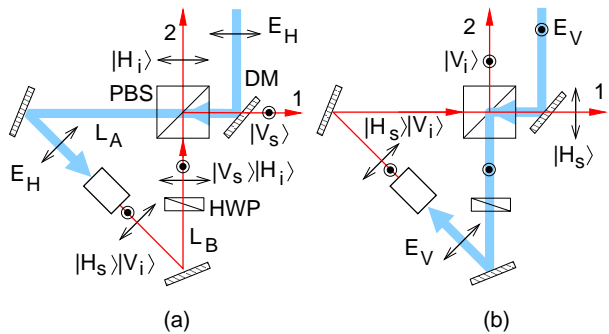}
  \end{center}
  \caption{Polarization Sagnac interferometer for type-II down-conversion for (a) $H$-polarized $E_H$ and (b) $V$-polarized $E_V$ pump components, with their counterclockwise and clockwise propagation geometry, respectively.  Orthogonally polarized outputs are separated at the PBS to yield polarization-entangled signal and idler photon pairs.  HWP: half-wave plate, DM: dichroic mirror, PBS: polarizing beam splitter.}
  \label{FIGURE_CIRCULAR}
\end{figure}

Figure~\ref{FIGURE_CIRCULAR} illustrates the workings of the bidirectionally pumped Sagnac down-conversion source.  Consider a classical pump field at frequency $\omega_p$ given by
\begin{equation}
\vec{E}_p = E_H \hat{e}_H + e^{i\phi_p} E_V \hat{e}_V\,,
\end{equation} 
where $\hat{e}_H$ and $\hat{e}_V$ are the horizontal ($H$) and vertical ($V$) polarization unit vectors, respectively, and $\phi_p$ is the relative phase between the $H$ and $V$ components.  The PSI is formed with a polarizing beam splitter (PBS), two high reflectors, a type-II phase-matched nonlinear crystal, and a half-wave plate (HWP).  The pump is directed into the PSI with a dichroic mirror (DM) that is highly reflective for the pump and highly transmissive for the signal (idler) output at $\omega_s$ ($\omega_i$).  We assume that the optical components of the PSI are appropriately designed for all three interacting wavelengths.  A technical design challenge is the triple-wavelength PBS and HWP\@.  In our experimental realization we have chosen to operate near degeneracy, $\omega_s \approx \omega_i$, so that dual-wavelength PBS and HWP can be readily obtained.  

As shown in Fig.~1(a), the $H$-polarized pump component traverses the interferometer counterclockwise and generates at the crystal an (unnormalized) output state given by $\exp(ik_p L_A) E_H |H_s\rangle |V_i\rangle$, which represents a collinearly propagating pair of $H$-polarized signal photon and $V$-polarized idler photon.  It is well known that in SPDC the sum of the output phases equals the pump phase \cite{Graham}, and hence the counterclockwise propagating output carries the phase picked up by the pump as it travels a length $L_A$ from the PBS to the crystal.  This output undergoes a $\pi/2$ polarization rotation at the HWP and the output state at the PBS is given by 
\begin{equation}
|\Psi_H\rangle = e^{i[k_p L_A + (k_s + k_i)L_B + \theta_s + \theta_i]}\, \eta_H E_H |V_s\rangle_1 |H_i\rangle_2\,.
\end{equation} 
The subscripts 1 and 2 refer to the two output ports labeled in Fig.~1 indicating that the signal and idler photons are separated at the PBS\@. The phases $\theta_s$ and $\theta_i$ are acquired by the signal and idler outputs when they pass through the HWP\@, and $\eta_H$ is the generation efficiency including propagation and absorption losses.

The $V$-polarized pump component follows the clockwise path in the PSI shown in Fig.~1(b).  For proper phase matching in the nonlinear crystal, the $V$-polarized pump is rotated by $\pi/2$ by the HWP and acquires a phase $\theta_p$.  The unnormalized output state at the PBS generated from the $V$-polarized pump component is
\begin{equation}
|\Psi_V\rangle = e^{i[\phi_p + k_p L_B + \theta_p + (k_s + k_i)L_A]}\, \eta_V E_V |H_s\rangle_1 |V_i\rangle_2\,.
\end{equation} 
Note that the relative pump phase $\phi_p$ shows up in $\Psi_V$ and the generation efficiency $\eta_V$ for the clockwise propagating outputs is not necessarily equal to that for the counterclockwise propagating ones in Eq.~(2).  

Recognizing that in free space $k_p = k_s + k_i$, the combined PSI output, to the lowest non-vacuum order, is the biphoton state
\begin{eqnarray}
\label{EntangledState}
|\Psi\rangle \propto  (|H_s\rangle_1 |V_i\rangle_2 + e^{i\phi}\beta |V_s\rangle_1 |H_i\rangle_2)\,,\\
\phi = \theta_s + \theta_i - \theta_p - \phi_p \,,\hspace{.25in}
\beta = \frac{\eta_H E_H}{\eta_V E_V}\,.
\end{eqnarray} 
The relative amplitude $\beta$ of the two bidirectionally pumped outputs can be adjusted to unity by changing the input pump component ratio $E_H / E_V$ to compensate for unequal generation efficiencies in the two directions, caused by, for example, different PBS transmission and reflection coefficients for the $H$ and $V$ polarizations.  The PSI is robust because, apart from an inconsequential overall phase factor, the relative phase $\phi$ between the two output components are independent of the path lengths $L_A$ and $L_B$ of the PSI\@. The term $\theta_s + \theta_i - \theta_p$ in $\phi$ is a function of the HWP's material dispersion and does not vary in time. Moreover, $\phi$ is fully adjustable by varying the pump phase $\phi_p$.  A singlet ($\phi = \pi$) or a triplet ($\phi = 0$) state can be easily generated by setting $\phi_p$ accordingly.  Therefore, the PSI down-conversion source is robust, phase stable, and adjustable, and, similar to other bidirectionally pumped sources \cite{MZIBiPump,PPLNBiPump}, it does not require spatial, spectral, or temporal filtering for the generation of polarization entanglement.

\begin{figure}
  \begin{center}
    \includegraphics[width=8cm]{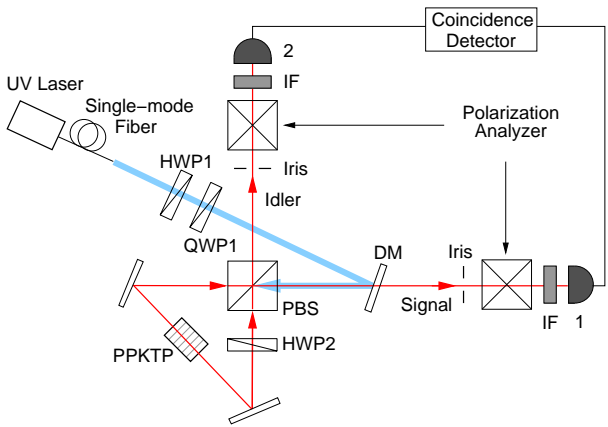}
  \end{center}
  \caption{Experimental setup for polarization Sagnac interferometer type-II down-conversion. PBS1 and QWP1 are used for adjusting the relative amplitude balance of the two counter-propagating down-conversion paths and the phase of the output biphoton state. HWP: half-wave plate, QWP: quarter-wave plate, DM: dichroic mirror, PBS: polarization beam splitter, IF: 1-nm interference filter centered at 810 nm.}
  \label{FIGURE_SETUP}
\end{figure}

Figure~\ref{FIGURE_SETUP} shows a schematic of the experimental setup.  A continuous-wave (cw) external-cavity ultraviolet (UV) diode laser at 405 nm was used as the pump source.  The linearly polarized UV pump laser was coupled into a single-mode fiber for spatial mode filtering and easy transport of the pump light to the PSI\@.    A half-wave plate (HWP1) and quarter-wave plate (QWP1) were utilized to transform the fiber-coupled pump light into the appropriate elliptically polarized light of Eq.~(1) to provide the power balance and phase control of the pump field ($\beta$ and $\phi$ in Eq.~(\ref{EntangledState})).  We typically used an input pump power of $\sim$3 mW.
  
We used a 10-mm-long (crystallographic $X$ axis), 2-mm-wide ($Y$ axis), 1-mm-thick ($Z$ axis) flux-grown PPKTP crystal with a grating period of 10.0 $\mu$m for frequency-degenerate type-II quasi-phase-matched collinear parametric
down-conversion.  The crystal was antireflection coated at 405 and 810 nm and temperature controlled using a thermoelectric cooler to within $\pm$0.01$^{\circ}$C\@.  The pump at $\sim$405 nm was weakly focused to a beam waist of $\sim$160 $\mu$m at the center of the crystal.  To fine tune the crystal temperature and pump wavelength combination, we measured the single-beam quantum interference in a setup similar to that reported in Ref.~\cite{SingleBeamQI}, obtaining a visibility of 98.9\% with a small aperture at the operating temperature of $28.87\pm 0.01^{\circ}$C and a pump wavelength of 404.96 nm.  

The PSI consisted of two flat mirrors that were highly reflecting at both wavelengths, a dual-wavelength half-wave plate (HWP2), and a dual-wavelength PBS that served as the input/output coupling element.  HWP2 was antireflection coated at the pump and output wavelengths and HWP2 worked well as a half-wave plate at both pump and output wavelengths.  However, the PBS was antireflection coated only at the output wavelength, and had good extinction ratio only at that wavelength. At the pump wavelength, it transmitted 73\% of the $H$ component and 3\% of the $H$ component leaked into the wrong output port. For the $V$ component of the pump, 80\% was reflected into the correct output of the PBS and 5\% was transmitted into the unwanted output port. For both polarizations, the rest of the pump power was either back reflected or absorbed by the PBS\@. The 10\% difference between the transmitted $H$-polarized component and the reflected $V$-polarized component was compensated for by adjusting the ratio $E_H /E_V$ using HWP1 and QWP1.   The unwanted pump components (transmitted $V$-polarized and reflected $H$-polarized components) did not contribute to down-conversion because they did not satisfy the phase-matching condition. 

At the PSI output, the polarization states of the spatially distinct signal and idler photons were analyzed with a combination of a HWP and a polarizer before detection by single-photon Si detectors (Perkin-Elmer SPCM-AQR-13).  We imposed a spectral bandwidth of 1 nm using an interference filter (IF) centered at 810 nm with a maximum transmission of 75\% at the center wavelength.  Two irises were used to control the acceptance angle of the detection system.  The outputs of the two single-photon detectors were counted and their coincidences were measured using a home-made coincidence detector \cite{Coincidence} with a 1-ns  coincidence window.  For singles count rate of $\sim$100\,000/s, we observed  accidental coincidence rate of $\sim$10/s.  In our data analysis, we subtract the accidental coincidence counts from the raw data.  

\begin{figure} 
  \begin{center}
          \includegraphics[width=8cm]{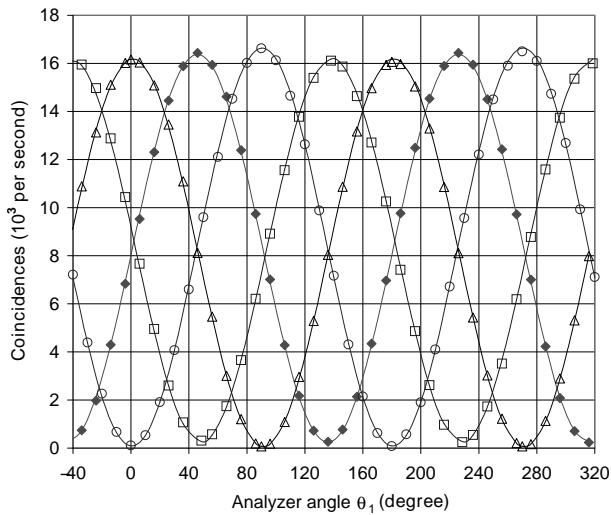}
  \end{center}
  \caption{Coincidence counts as a function of signal polarization analyzer angle $\theta_1$ for different settings of idler polarization analyzer angle $\theta_2$: 0$^\circ$(open circle), 46$^\circ$(open square), 90.5$^\circ$(open triangle), 135$^\circ$(solid diamond). The biphoton output was set to be a singlet state.  Solid lines are best sinusoidal fits to data.  Each data point was averaged over 40 s and the pump power was 3.28 mW.}
  \label{FIGURE_SINUSOID}
\end{figure}

Figure~\ref{FIGURE_SINUSOID} shows the coincidence counts as a function of the signal polarization analyzer angle $\theta_1$ when the idler polarization analyzer angle $\theta_2$ was fixed at four different values: 0$^\circ$, 46$^\circ$, 90.5$^\circ$, and 135$^\circ$. We used HWP1 and QWP1 to set the PSI output to be in a biphoton singlet state ($\beta = 1, \phi = \pi$).  The overlaid solid curves are best sinusoidal fits to the corresponding data.  For an input pump power of 3.28 mW (measured before DM), 1-nm IF, and a full divergence collection angle of 12.5 mrad, we measured a flux of polarization-entangled photon pairs of $\sim$5\,000 detected pairs/s/mW. Quantum-interference fringe visibility $V$ is defined by ($C_{\rm max} - C_{\rm min})/(C_{\rm max} + C_{\rm min}$),
where $C_{\rm max}$ is the maximum coincidence counts and $C_{\rm min}$ is the minimum coincidence counts.  We calculate the visibilities for the four sets of measurements in Fig.~\ref{FIGURE_SINUSOID} based on best-fit parameters: $V_0=99.09\pm0.07$, $V_{46}=96.85\pm0.12$, $V_{90.5}=99.32\pm0.04$, and $V_{135}= 97.18\pm0.09$. These high visibilities imply that the output state is very close to the ideal polarization-entangled biphoton singlet state. 

We have measured the S parameter in the Clauser-Horne-Shimony-Holt (CHSH) form of Bell's inequality \cite{CHSH}. By following the same method described in Ref.~\cite{NoncollinearEntangle,CHSHMeasure}, the expectation function $E$ for a set of ($\theta_1, \theta_2$) can be calculated by averaging the product of two measurements from paths 1 and 2. Each of the two measurements gives one of two possible results: $+1$ when the measured polarization is parallel to the polarization analyzer angle, and $-1$ when the polarization is perpendicular to that angle.  The product yields $+1$ if the two polarization measurements correspond to ($\theta_1, \theta_2$) or ($\theta_1+\pi/2, \theta_2+\pi/2$), and $-1$ if they are equal to ($\theta_1+\pi/2, \theta_2$) or ($\theta_1, \theta_2+\pi/2$). Ideally, we would use 4 single-photon detectors and simultaneously count the four coincidences for every generated pair.  However due to limitations in our setup, we could measure only one type of coincidences at one time.  Therefore, we construct $E$ from 4 types of coincidences using
\begin{equation}
E(\theta_1, \theta_2)=\frac{C_{++}-C_{+-}-C_{-+}+C_{--}}{C_{++}+C_{+-}+ C_{-+}+C_{--}}\,.
\end{equation}
$C_{+-}$ is the coincidence count for analyzer 1 set at $\theta_1$ and analyzer 2 set at $\theta_2+\pi/2$, and similar definitions are made for  $C_{++},C_{-+},C_{--}$.  
The parameter 
\begin{equation}
S=|E(0,\frac{7\pi}{8})+E(-\frac{\pi}{4},\frac{7\pi}{8})+E(-\frac{\pi}{4},\frac{5\pi}{8})-E(0,\frac{5\pi}{8})|
\end{equation}
requires four different $E$ measurements, totaling 16 sets of coincidence measurements.  Each measurement set took 40 s to complete and we obtain $S = 2.7645\pm0.0013$, which corresponds to the violation of the classical limit of 2 by more than 500 standard deviations.

\begin{figure}
\begin{center}
\includegraphics[width=8cm]{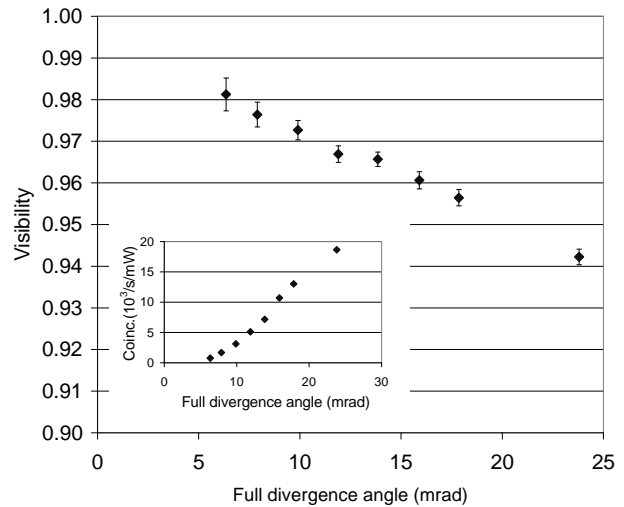}
\end{center}
  \caption{Plot of quantum-interference visibility as a function of full divergence angle for $\theta_2=45^\circ$. Inset: measured flux of polarization-entangled photon pairs versus full divergence angle.}
  \label{FIGURE_VISIBILITY}
\end{figure}

We have examined the characteristics of the PSI down-conversion source at different collection divergence angles.  Figure~\ref{FIGURE_VISIBILITY} shows the visibility and flux of the generated polarization-entangled photon pairs as we varied the size of the collection irises, showing reduced flux but higher visibility for smaller divergence angles.  We attribute the dependence of the visibility on the iris size to wavefront distortion of the PBS used in the PSI\@.  The commercially available PBS was made from two coated prisms that were cemented together at the hypotenuse side with a wavefront distortion of a quarter of a wavelength at 633 nm.  This specification is worse than intracavity optics that are typically specified at a tenth of a wavelength at 633 nm.  The wavefront distortion is less when the collection divergence angle is small, thus resulting in higher visibilities for smaller apertures, as observed in Fig.~4. 

 To clarify this wavefront issue, we have measured the classical fringe visibility of the PSI using the pump light as the probe.  To observe the classical fringes we oriented HWP2 in Fig.~2 so that its fast axis was parallel to the $H$ polarization.  We observed a similar divergence-angle dependence of the classical visibility as in Fig.~4, suggesting that the wavefront is not constant over a large area.  This could also be confirmed when we displaced the center of the iris slightly off axis ($\sim$20 $\mu$m), we observed a small drop in the classical visibility which could then be recovered by adjusting the pump phase $\phi_p$. 

It is useful to compare the present results with similar measurements obtained in our previous bidirectionally pumped MZ source of Ref.~\cite{MZIBiPump}, with a reported spectral brightness of 4\,000 pairs/s/mW/nm at a 90\% visibility.  In contrast, the measured flux (full divergence angle of 12.5 mrad) from the PSI source is 5\,000 detected pairs/s/mW in a 1-nm bandwidth at a 96.8\% visibility, which is higher than the MZ source by 25\% and with a higher visibility.  We attribute the improvement to a more compact size of the PSI setup that not only improves the collection efficiency (higher flux) but also minimizes spatial mode clipping inside the interferometer (higher visibility).  Under ideal conditions, the visibility should be constant and near unity for different aperture sizes.  However, a small amount of spatial mode distinguishability remained due to wavefront distortion caused by the PBS of the polarization Sagnac interferometer.  We note that without any iris, we obtained a flux of 22\,750 pairs/s/mW with a still-high visibility of 93.0\%.  We should point out that the low extinction ratio of the PBS at the pump wavelength reduces the useful pump power to 76\% of that measured after QWP1 (see Fig.~2).  A higher-quality PBS would have yielded a $\sim$32\% higher flux, but we have not included this type of correction in the presently reported values.

In conclusion, we have demonstrated a compact polarization-entangled photon source using a polarization Sagnac interferometer with improved stability, higher flux, and higher visibility than previous type-II phase-matched sources.  The Sagnac geometry eliminates the need for stabilizing the interferometer and allows the biphoton output state to be easily and precisely controlled.  This source can also be configured to produce hyperentangled photon pairs, with entanglement in both momentum and polarization, that can be potentially useful for implementing single-photon two-qubit quantum logic \cite{CNOT,SWAP}.

This work was supported by the Department of Defense Multidisciplinary University Research Initiative program under Army Research Office grant DAAD-19-00-1-0177, by MIT Lincoln Laboratory, and by ARDA through a NIST grant 60NANB5D1004.

\end{document}